\documentclass[reprint, twocolumn,
 amsmath,amssymb,aps,
 prl,
]{revtex4-2}

\usepackage{graphicx}
\usepackage{dcolumn}
\usepackage{bm}
\usepackage{physics}
\usepackage{xcolor}

\begin{document}

\title{Orbital dynamics in 2D topological and Chern insulators}

\author{Daniel Fa\'ilde }

\affiliation{Departamento de Física Aplicada, Instituto de Investigacións Tecnolóxicas,Universidade de Santiago de Compostela,E-15782 Campus Vida s/n, Santiago de Compostela, Spain}

\author{Daniel Baldomir}

\affiliation{Departamento de Física Aplicada, Instituto de Investigacións Tecnolóxicas,Universidade de Santiago de Compostela,E-15782 Campus Vida s/n, Santiago de Compostela, Spain}

\date{\today}

\begin{abstract}
Within a relativistic quantum formalism we examine the role of second-order corrections caused by the application of magnetic fields in two-dimensional topological and Chern insulators. This allows to reach analytical expressions for the change of the Berry curvature, orbital magnetic moment, density of states and energy determining their canonical grand potential and transport properties. The present corrections, which become relevant at relatively low fields due to the small gap characterizing these systems, unveil a zero-field diamagnetic susceptibility which can be tuned by the external magnetic field.

\end{abstract}

\maketitle

One of the most special features of the topological insulators (TIs) is the presence of protected helical states on their boundaries which are responsible for their singular transport properties \cite{zhang2009topological,Bernevig1757}. Just as their robustness against non-magnetic impurities or external fields, the quantization of their transport properties also depend directly on the topology by means of a topological invariant which can be defined according to the intrinsic symmetries of the system and its dimensionality \cite{PhysRevLett.95.146802,PhysRevB.76.045302,PhysRevLett.97.036808,PhysRevB.78.195125}. In time-reversal symmetry broken systems as well as in two-dimensional topological insulators this invariant is the first Chern number $C$ obtained throughout the integral of the Berry curvature over the momentum space \cite{PhysRevB.78.195424,PhysRevB.74.235111,PhysRevB.81.115407}. Besides the well-known relation between the electric conductivity and polarization with the topological invariant \cite{Konig766,RevModPhys.82.1959,PhysRevB.47.1651}, great and original advances have been done to address the thermoelectric response of systems with non-zero Berry curvature in presence of electric and magnetic fields \cite{PhysRevLett.95.137204,PhysRevLett.97.026603,PhysRevLett.107.236601,PhysRevLett.95.137205,PhysRevLett.99.197202,zhang2016berry}. These studies take the semi-classical equations of motion for the Bloch electrons or a  non-relativistic quantum formalism to derive magnetization and electric and thermal currents for a wide variety of compounds. These are the bases used to study planar Hall and chiral anomaly effects in topological insulators and Weyl semimetals through Boltzmann transport equation with in-plane magnetic fields \cite{Nandy2018,PhysRevLett.119.176804}.

Recently, the original studies have been extended by  addressing second-order corrections through the Lagrangian formalism \cite{PhysRevLett.112.166601,PhysRevB.91.214405}. However, determining these quantities in a purely quantum way for the special case of topological insulators and Chern insulators, which present a non-trivial Berry curvature, involves some difficulties. First, we have to deal with a relativistic system described through a Dirac Hamiltonian \cite{PhysRevLett.101.246807,PhysRevB.81.115407}, where spin and angular momentum are no longer good quantum numbers of the system and the velocity  differs from the momentum as in their usual non-relativistic form $v=p/m$. Secondly, the evolution of eigenstates needs to be considered adiabatically; i.e., keeping the final and initial states of the system the same along the perturbation to preserve Berry phase effects. This leads us to treat with gauge dependent and divergent corrections to the system eigenstates that are identified and removed to get the usual equations of motion for non-zero Berry curvature systems but now 
in the proper relativistic context of these materials at low energies.

With this approach, we give analytical expressions to show how the introduction of a perpendicular magnetic field in 2DTIs and Chern insulators produces a modulation of the Berry curvature, which can affect its shape dramatically, but keeping the Chern number $C$ of the system invariant. This effect is independent of the magnitude and time dependence of the magnetic field $B$ at least until adiabaticity is lost or other effects such as the Zeeman splitting need to be considered. Behind these results, we can find the additional contributions to the density of states, orbital magnetic moment and energy corresponding to second-order corrections in perturbation theory. These terms must be taken into account at relatively low external magnetic fields due to the small topological gap characterizing these systems. In particular, we show that for the energy only those terms coming from the modified orbital magnetic moment, which are associated with the correction to the Berry potential, are necessary, while the other obtained with the semi-classical Lagrangian formalism in a relativistic particle-hole symmetric system vanish \cite{PhysRevB.91.214405}. Additionally, we observe a modified density of states that is strongly sensitive to the sign of the magnetic field and whose dispersion differs substantially from its first-order expansion \cite{PhysRevLett.95.137204}. These results can be directly introduced to determine explicitly the thermodynamic grand potential and hence the transport magnitudes in such systems, or in the Dirac oscillator Hamiltonian, as an argument to demonstrate how certain type of chiral photons or phonons can couple to the topological electrons preserving their topology and  time-reversal symmetry $\hat{T}$ necessary for the presence of Kramer's pairs \cite{moshinsky1989dirac,PhysRevA.77.063815,PhysRevA.76.041801,failde2020emergent}.

The quantum-materials version of the Dirac equation substitutes the light velocity $c$ of the particle by the Fermi velocity of the electrons, as well as in some cases it incorporates a momentum dependence in the mass associated with the k-dependent energy dispersion \cite{shen2012topological},
\begin{equation}
    i\hbar v_F \gamma^\mu \partial_\mu \Psi+m(\boldsymbol{k}) v_F^2 \Psi=0
\end{equation}
where $\gamma^\mu$ are the gamma matrices, $\mu=1,2,3,4$ and $\partial_\mu$ is the 4-gradient. In two-dimensional systems, where the term proportional to $p_z$ disappears, the Dirac Hamiltonian can be decoupled into two time-reversal symmetry-related copies of a two-level Dirac Hamiltonian which is appropriate to introduce us to the non-trivial topological formalism \cite{PhysRevB.81.115407,Shan_2010}.
\begin{equation}
    H= \left(\begin{array}{cc} M({\bf k}) & \hbar v_F k_-\\ \hbar v_F k_+ & - M({\bf k}) \end{array}\right)
\label{Hamiltonian}
\end{equation}
Here $v_F$ is the Fermi velocity, $\hbar$ is the reduced Planck constant, $k_\pm=k_x\pm ik_y$ and $k=\sqrt{k_x^2+k_y^2}$. The term $M(\boldsymbol{k})=M-\mathcal{B}k^2$, representing the gap ($2M$) in the center of the Brillouin zone and its parabolic dependence, breaks the time-reversal symmetry of the system allowing a suitable characterization of the topology by means of the topological invariant Chern number $C$ derived from the integral of the Berry curvature; i.e., $C=1/(2\pi)\int \boldsymbol{\Omega}^n(k) d\boldsymbol{k}$ being $\boldsymbol{\Omega}^n=-2Im \bra{\partial_{k_x}n}\ket{\partial_{k_y} n} \boldsymbol{\hat{z}}$ the Berry curvature of the eigenstate $n$. As it is known, to get a non-zero Chern number $C=\pm 1$, $M$ and $\mathcal{B}$ must have the same relative signs ($M\mathcal{B}>0$), implying that the incorporation of a spin-orbit coupling gets crucial to produce the crossing between the bands that precede the non-trivial topological regime \cite{Konig766,PhysRevB.81.115407}. The introduction of a magnetic field $\boldsymbol{B}=(0,0,B)$ in the z-direction breaks the translational symmetry in $x$ and $y$ directions, which is evident by choosing an axial gauge $\boldsymbol{A}=(-By/2,Bx/2,0)$ to enter the perturbation in the Hamiltonian through the Peierls substitution $\boldsymbol{p} \rightarrow \boldsymbol{p}+e\boldsymbol{A}$, being $-e$ the electron charge. In such situation, the correction to the eigenstates by a perturbation, which corrects the particle momentum, has the following form up to first order \cite{shen2012topological}
\begin{equation}
    \ket{n} \rightarrow \ket{n} -i\hbar \sum_j\frac{\bra{m}\partial k_j/\partial t\ket{\partial_{k_j} n}}{\xi_n -\xi_m} \ket{m}
\end{equation}
being $i$ the imaginary number, $j=x,y$ denoting the spatial coordinates and $\ket{n}$ and $\ket{m}$ the eigenstates of the system. Let's label $\ket{+}$ and $\ket{-}$ the eigenstates with energy $\xi^\pm=\pm \sqrt{M(k)^2+\hbar^2v_F^2k^2}$ of Hamiltonian (2). The presence of a field $\boldsymbol{B}$ implies the existence of a Lorentz force in the system which for the $x$ and $y$ directions is $\hbar \frac{\partial k_x}{\partial t}= -\frac{e}{2} \frac{\partial}{\partial t}(By)$ and $\hbar \frac{\partial k_y}{\partial t}= \frac{e}{2} \frac{\partial}{\partial t}(Bx)$. Thus, considering a constant uniform $B$, we have for the positive eigenstates that corrections result in the following form

\begin{eqnarray}
    \nonumber \ket{+} \rightarrow \ket{+} +i\frac{e B}{2}  \frac{\bra{-} \hat{v}_y \ket{\partial_{k_x} +}}{2 \xi} \ket{-}\\
    -i\frac{e B}{2} \frac{\bra{-}\hat{v}_x\ket{\partial_{k_y} +}}{2 \xi} \ket{-}
\label{corrections}
\end{eqnarray}
where $\hat{v}_j=i/\hbar[\hat{H},\hat{r}_j]=\hbar^{-1} \partial_{k_j} H=v_F \sigma_j-2\mathcal{B}k_j\sigma_z$ is the velocity operator in the $j$ direction, we have taken $\xi_+ -\xi_-=2\xi$ provided that $H$ is particle-hole symmetric and where the system eigenstates can be found to be 

\begin{equation}
    \ket{+}=\frac{1}{\sqrt{2}}\left[\begin{array}{c} \sqrt{1+\frac{M({\bf k})}{\xi}} \\ e^{i\phi} \sqrt{1-\frac{M({\bf k})}{\xi}} \end{array}\right]
\end{equation}
\begin{equation}
    \ket{-}=\frac{1}{\sqrt{2}}\left[\begin{array}{c} \sqrt{1-\frac{M({\bf k})}{\xi}} \\ -e^{i\phi} \sqrt{1+\frac{M({\bf k})}{\xi}} \end{array}\right]\\[10pt]
\end{equation}
being $\phi=arctan(k_y/k_x)$. For simplicity, we proceed by setting the Hamiltonian parameter $\mathcal{B}$ as zero. As it seems logical, it is worthy to note that the corrections in Eq.\,\eqref{corrections} are proportional to the product of the magnetic field with the $z$-component of orbital magnetic moment $\hat{m}_z=-e/2 (\hat{x}\hat{v}_y-\hat{y}\hat{v}_x)$ of the Bloch electrons \cite{PhysRevLett.95.137205,PhysRevB.59.14915,Chang_2008}.
\begin{equation}
   \ket{+}\rightarrow \ket{+}+\frac{eB}{2}\frac{\bra{-}\boldsymbol{\hat{r} \times \hat{v}}\ket{+}}{2\xi} \ket{-}
\end{equation}
However, in order to get a proper definition of the angular momentum and orbital magnetic moment on the band $n$, the previous expression needs to be corrected by $\boldsymbol{m}=-e/2 \left( \boldsymbol{r}\times (\boldsymbol{v}-\langle \boldsymbol{v^n}\rangle)\right)$, where $\langle v^n\rangle=\bra{n}v^n\ket{n}=\hbar^{-1} \partial_k \xi^n$ is the average velocity of the electrons in band $n$. This is equivalent to the addition of the center-of-mass position $r_c$ and its velocity in the Lagrangian formalism \cite{PhysRevLett.112.166601,PhysRevB.91.214405}. In this way, we can define properly the orbital magnetic moment \cite{PhysRevLett.95.137204,PhysRevLett.97.026603,PhysRevLett.95.137205,Chang_2008},
\begin{equation}
    m^n(\boldsymbol{k})=-i \frac{e}{2\hbar} \bra{\nabla_{\boldsymbol{k}} n} \times (H-\xi^n) \ket{\nabla_{\boldsymbol{k}} n}
\end{equation}
which results to be $m_z^n=\hbar^{-1}e \xi^n \Omega^n$ for a two-dimensional system as Eq.\,\eqref{Hamiltonian}, and the first-order corrections to the energy $\xi^n_1=-\boldsymbol{m}\cdot\boldsymbol{B}$. Nevertheless, the difficulties arise in Eq.\,\eqref{corrections}  when one computes the matrix elements 
\begin{equation}
      \bra{-} \left( \begin{array}{cc} 0 & 1 \\ -1 & 0 \end{array} \right)  \ket{\partial_{k_x} +} + \bra{-} \left( \begin{array}{cc} 0 & -i \\ -i & 0 \end{array} \right)  \ket{\partial_{k_y} +}=\frac{1}{2k}
\label{divergence}
\end{equation}
where it appears a divergence at zero particle momentum after gauge dependent terms have been removed. This behaviour is also present when computing velocity corrections and hence this contribution must be unphysical given that the force exerted by a magnetic field on a particle at rest is zero. We can solve this problem by decoupling the different contributions produced by the perturbation through the other definition of the velocity operator $\hbar^{-1} \partial_k H$. In this way, we can identify the ill-defined terms and properly obtain the corrections for the electron\textquotesingle s velocity in topological systems. Rewriting Eq.\,\eqref{corrections} by using that $\bra{m} \partial_{k_j} H\ket{\partial_{k_l}n}=\partial_{k_j}(H\ket{m})^*\ket{\partial_{k_l}n}-\bra{\partial_{k_j}m}H\ket{\partial_{k_l}n}$ 
\begin{widetext}
\begin{equation}
\begin{split}
    \ket{+}\rightarrow \ket{+}  +i \frac{eB}{4\xi} \Bigg[ &  \left( \frac{1}{\hbar} \partial_{k_y}\xi^-\bra{-}\ket{\partial_{k_x}+}- \frac{1}{\hbar} \partial_{k_x}\xi^-\bra{-}\ket{\partial_{k_y}+} \right)
    + \frac{\xi^-}{\hbar} \bigg( \bra{\partial_{k_y}-}\ket{\partial_{k_x}+}- \bra{\partial_{k_x}-}\ket{\partial_{k_y}+}\bigg) \\
    & - \frac{1}{\hbar} \bigg( \bra{\partial_{k_y}-}H\ket{\partial_{k_x}+}- \bra{\partial_{k_x}-}H\ket{\partial_{k_y}+}\bigg) \Bigg]  \ket{-}
\end{split}
\label{Eq9}
\end{equation}
it can be shown that the third term is purely gauge dependent by rotations $e^{i\phi}$ of the eigenstates, i.e. for $\ket{n'}=e^{-i\phi}\ket{n}$ and $\ket{m'}=e^{-i\phi}\ket{m}$ it changes its sign, and thus we can set one in which this term goes to zero. On the other hand, the first and second terms give a contribution equal to $-\frac{eB\Omega^+}{4\hbar} \frac{\hbar v_F k}{M}$ and $-\frac{eB\Omega^+}{4\hbar} \frac{M}{\hbar v_F k}$ respectively, being $\Omega^+=- \frac{\hbar^2v_F^2 M}{2 \xi^3}$ the Berry curvature of the conduction band of Hamiltonian (2) and leading their sum to Eq.\,\eqref{divergence} after rearranging terms.
\end{widetext}

Working with free divergent terms, i.e. the first, which must be considered twice due to the redefinition of the orbital magnetic moment, we can now easily compute velocity corrections in both directions. In fact, it is straightforward to see that corrections due to transverse components disappear and only longitudinal terms remain. Thus, we obtain the following corrections to the velocity which apply to both conduction and valence band by substituting their associated energy and curvature,
\begin{eqnarray}
    v^n_{j}\rightarrow \frac{1}{\hbar} \partial_{k_{j}}\xi^n + \frac{1}{\hbar} \partial_{k_{j}}\xi^n\frac{e\boldsymbol{B \cdot \Omega^n}}{\hbar}+\mathcal{O}(B^2)
    \label{positive}
\end{eqnarray}
where $v_j^n=\bra{n}\hat{v}_j\ket{n}$ the average velocity in the band $n$ for the component $j$ and $\mathcal{O}(B^2)=-1/(4\hbar)\, \partial_{k_j}\xi^n\,(eB\Omega/\hbar)^2\hbar^2v_F^2k^2/M^2$ second-order corrections. In a simple way, we are observing the coupling effects between the magnetic field and the Berry curvature, which can be viewed like a magnetic field in the k-space on each band of Hamiltonian (2). Thus, introducing a perpendicular $\boldsymbol{B}$ in these systems enhances or decreases the \textit{field} felt by their electrons depending on the relative sign between $\boldsymbol{B}$ and $\boldsymbol{\Omega}$. For instance, the conduction band of Hamiltonian (2) for $M<0$ has a positive Berry curvature in the $z$ direction and therefore an opposite magnetic field will decrease the velocity of their electrons and the Berry curvature even doing it zero or changing its sign. Given that the Lorentz force is radial this process causes an accommodation of the charge without involving any net current, as it can be checked by computing the integral of the previous expression. This is intrinsically related to the renormalization process affecting the phase-space volume and density of states for non-zero Berry curvature systems as we are going to show \cite{PhysRevLett.95.137204,PhysRevLett.97.026603,Duval, PhysRevLett.112.166601}.

Complementing these effects, we can also consider contributions associated with a slow time dependence for $B$ which incorporates a transverse term that can be easily transformed through Faraday's law into the well-known anomalous velocity using that $E_x=\frac{1}{2}\frac{\partial B}{\partial t} y$ and $E_y=-\frac{1}{2}\frac{\partial B}{\partial t}x$. The obtained expression up to first-order
\begin{equation}
    v_j^n\rightarrow v_j^n \left(1+\frac{e \boldsymbol{B \cdot \Omega}}{\hbar} \right) + \frac{e}{\hbar} (\boldsymbol{E \times \Omega^n})_j
\end{equation}
represents the velocity of the electrons in the band $n$ of a Chern insulator Eq.\,\eqref{Hamiltonian} or in one of the two branches of a two-dimensional topological insulator in a slowly variant time-dependent magnetic field. In contrast to the first contribution, the second term is associated with the electromotive force $\mathcal{E}$ generated by the variation of $B$ which couples to the Berry curvature to produce a transverse and non-zero electric current. 

Setting aside this latter case, we wondered, as we postulated before if one of the crucial magnitudes for the topology and the transport, the Berry curvature, has experimented changes under this procedure. For the calculation it is convenient to employ an axial gauge $\boldsymbol{A}=(-By/2,Bx/2,0)$ from which, as we showed, we are able to write the correction to the eigenstates in an easy to handle form
\begin{equation}
    \ket{+} \rightarrow \ket{+} -\frac{e \boldsymbol{B \cdot \Omega^+}}{2 \hbar} \frac{\hbar v_F k}{M} \ket{-}\\
    \label{curvature corrections}
\end{equation}
Once we formulated the correction of the eigenstates the calculation of the Berry curvature corrections for the conduction band can be achieved by applying $-2Im \bra{\partial_{k_x}+}\ket{\partial_{k_y}+}$ or $\partial_{k_x}\mathcal{A}_y-\partial_{k_y}\mathcal{A}_x$ in Eq.\,\eqref{curvature corrections}, being $\mathcal{A}_i=i\bra{+}\ket{\partial_{k_i} +}$ the Berry potential and  $\ket{+}$ the modified eigenstate. In fact, it is straightforward to show that the obtained corrections to the Berry potential are the same as the theoretically presented in ref.\;\cite{PhysRevLett.112.166601}. After some algebra, it can be proved that Berry curvature turns out in the following form
\begin{equation}
    \Omega^+ \rightarrow \Omega^+ \left(1+2\frac{e \boldsymbol{B \cdot \Omega^+}}{\hbar}\right)-2 \Omega^+ \frac{e \boldsymbol{B \cdot \Omega^+}}{\hbar} \frac{\hbar^2 v_F^2 k^2}{M^2}
    \label{final corrections}
\end{equation}
demonstrating how a perpendicular magnetic field $\boldsymbol{B}$ modulates the Berry curvature and the field seen by the electrons in these topological systems. Besides the familiar first term in Eq.\,\eqref{final corrections} we have obtained a second contribution in the corrections which affects the Berry curvature at $k$ out of $k=0$. This term is important at intermediate values whereas it falls to zero when $k\rightarrow \infty$ and $k=0$, although it can be shown to be tuned and even to disappear if we consider some energy dependence in the field $B$. 

\begin{figure}[t]
\includegraphics[scale=0.9356]{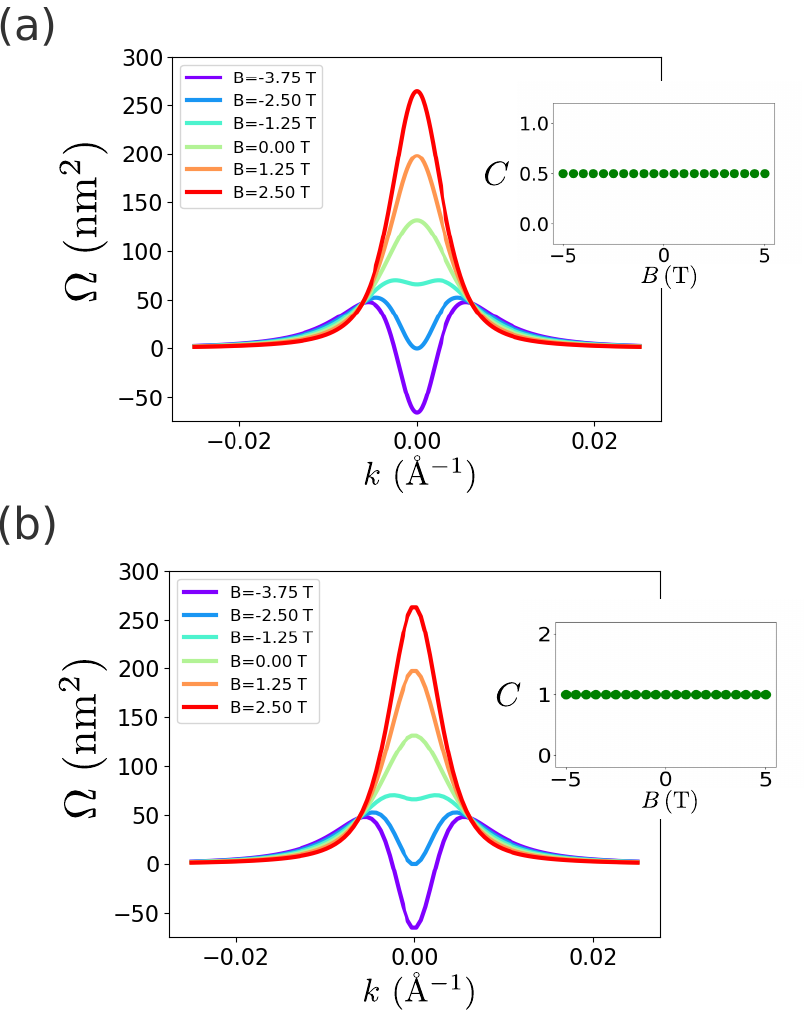}
\caption{Berry curvature corrections for the conduction band of Hamiltonian (2) for different values of $\boldsymbol{B}=(0,0,B)$. The inset shows the numerical calculations of the Chern number $C=1/(2\pi) \int \boldsymbol{\Omega} d\boldsymbol{k}$. Plot parameters are (a) $M=-0.025$ eV, $v_F=6.17\cdot$10$^5$m/s, $\mathcal{B}=0.0$ eV\AA$^2$  and (b) $\mathcal{B}=-5.0$ eV\AA$^2$}
\label{B=0}
\end{figure}

\begin{figure}[t]
\includegraphics[scale=0.8]{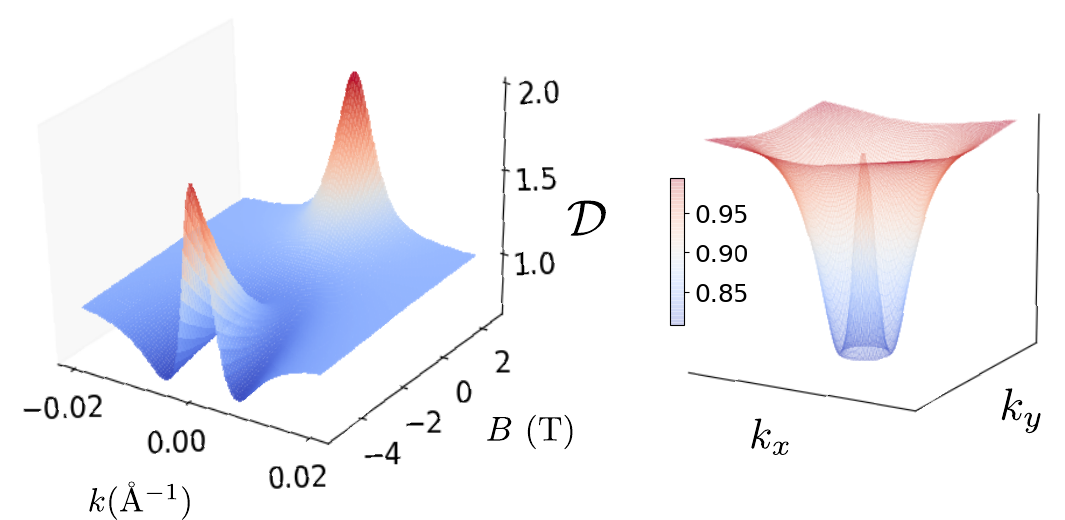}
\caption{Modified density of states of the conduction band as a function of $k$ and $B$. Right panel shows the momentum dependence of $\mathcal{D}$ for the particular case $B=-m^2v_F^2/(e\hbar)$.}
\label{Density}
\end{figure}

Since the Berry curvature has been modified, the next step is to compute the first Chern number C given its relation to the transport and hence with different physical observables. With this purpose, we can consider a uniform magnetic field of the form $B \propto m_e^2 v_F^2/(\hbar e)$ just like in ref. \cite{failde2020emergent}, where the translation of the Berry curvature into a real field $b$ was made using the magnetic flux quantization of helical orbits in terms of the Chern. As it has been analyzed, this field is closely related to the critical field $B_c$ needed to create electron-hole Schwinger pairs in the vacuum. However, this consideration is not necessary and one can also proceed equally by extracting $B$ from the integral and computing it numerically (Fig.\;\ref{B=0}(a)). Choosing the first option, the term $2e\boldsymbol{B \cdot \Omega}/\hbar$ can be written as $-M^3/\xi^3$ given that $M=m_ev_F^2$ and hence
\begin{equation}
    C= \frac{1}{2\pi} \int \boldsymbol{\Omega} d \boldsymbol{k} \rightarrow \frac{1}{2\pi}  \int \boldsymbol{\Omega} \left( 1-2\frac{M^3}{\xi^3}+\frac{M}{\xi}  \right) d\boldsymbol{k}
\end{equation}
where $d\boldsymbol{k}=2\pi k dk$. By using that $\Omega^{\pm}=\pm \partial/\partial k^2 (M/\xi)$ it is straightforward to see that the sum of second and third terms in the integral cancel
\begin{equation}
\begin{split}
    \frac{1}{2} &\left[\int_0^\infty \frac{1}{ 2}\frac{\partial}{\partial k^2} \left( \frac{M}{\xi} \right)^2 dk^2  \right. \\ 
    &-\left. \int_0^\infty \frac{1}{ 2}\frac{\partial}{\partial k^2} \left( \frac{M}{\xi} \right)^4 dk^2  \right]=\frac{1}{4} \frac{M^2}{\xi^2} \bigg\rvert_0^\infty-\frac{1}{4} \frac{M^4}{\xi^4} \bigg\rvert_0^\infty=0
\end{split}
\end{equation}
As consequence, the Chern number of the band does not change even though the Berry curvature does it. This occurs independently of the magnitude and time dependence of $B$ until higher-order effects need to be considered or adiabaticity is lost and it is consistent with the preservation of quantized conductivities in the quantum Hall regime. These calculations can also be derived for non-zero but small $\mathcal{B}$ values ($v_F^2>>2BM/\hbar^2$). In this case, after neglecting terms in the energy derivative $\partial_{k_j}\xi$ in Eq.\,\eqref{Eq9}, the curvature corrections turn out into a more tedious expression
\begin{equation}
\begin{split}
    \Omega^+ &\rightarrow \Omega^+ \left(1+2\frac{e\boldsymbol{B\cdot \Omega^+}}{\hbar} M \frac{M-\mathcal{B}k^2}{(M+\mathcal{B}k^2)^2} \right)\\
    \medskip
    &-2\Omega^+ \frac{e \boldsymbol{B \cdot \Omega^+}}{\hbar} \frac{\hbar^2 v_F^2 k^2}{(M+\mathcal{B}k^2)^2} \left(1-3\mathcal{B}\frac{M-\mathcal{B}k^2}{\hbar^2v_F^2} \right)
\end{split}
\end{equation}
but for which the Chern number $C$ is constant and well-defined by an integer value, i.e. $\pm 1$ if $M\mathcal{B}>0$ and $0$ if $M\mathcal{B}<0$ (Fig.\;\ref{B=0}(b)). Notice that here $\Omega^+=-\hbar^2v_F^2(M+\mathcal{B}k^2)/(2\xi^3)$. In both cases, there is a value ($B\approx-2.5$T for the values of $M$ and $v_F$ taken) for which the Berry curvature falls to 0 at the $\Gamma$ point. This value is not other than the one delimited by the equation $b=2m_e^2v_F^2/(\hbar e)$ in ref. \cite{failde2020emergent} with a difference of a factor 1/2 which comes from the redefinition of the orbital magnetic moment. This opens the possibility to enter in a regime where electron-hole pair creation might be experimentally accessible for certain $k$ values. In contrast, we find that the case with $eB\Omega/\hbar=-1$ making zero the density of states $\mathcal{D}$ which arises when considering constant the Berry curvature \cite{PhysRevLett.95.137204}, actually does not take place for $k=0$. For these values of $B$, second-order corrections need to be taken into account and the density of states writes as $\mathcal{D}=1+eB\Omega^*/\hbar$ with $\Omega^*$ the modified Berry curvature displayed in Eq. (14) or (16) \cite{PhysRevLett.112.166601,PhysRevB.91.214405}. This function has a minimum at $k\neq0$ (Fig.\;\ref{Density}) which can be tuned by $B$ becoming zero for sufficiently high magnetic fields.

Furthermore, we are also in position to write second-order corrections to the energy given that the matrix element $\bra{-}\Delta H\ket{+}=-B \bra{-}m_z\ket{+}$ has been computed before. Then, we directly obtain that
\begin{equation}
\xi\rightarrow \xi - \boldsymbol{m \cdot B} + \frac{1}{2} \frac{(\boldsymbol{m \cdot B)}^2}{\xi}\frac{\hbar^2v_F^2k^2}{M^2}
\label{energy}
\end{equation}
where $-\boldsymbol{m \cdot B}$ is the well-known first-order response and the third term comes from second-order effects. This formula seems to enter in conflict with the one obtained from a semi-classical Lagrangian theory \cite{PhysRevB.91.214405}, in which the energy up to second-order for a relativistic particle-hole symmetric system as Eq.\;\eqref{Hamiltonian} is
\begin{equation}
\begin{split}
\bar\xi=\xi_0&-\boldsymbol{m \cdot B}+\frac{1}{4}\boldsymbol{m \cdot B} \frac{e\boldsymbol{B \cdot \Omega}}{\hbar} \\
&-\frac{1}{8}e^2\epsilon_{sik}\epsilon_{tjl}B_s B_t g_{ij}\alpha_{kl}-eB (\mathcal{A^*}\times v_0)
\end{split}
\end{equation}
where $g_{ij}=Re \bra{\partial_i n}\ket{\partial_j n}-\bra{\partial_in}\ket{n}\bra{n}\ket{\partial_j n}$ is the quantum metric in the k-space, $\alpha_{kl}=\partial_{kl}\xi_0/\hbar^2$ the inverse of the effective mass tensor, $v_0=\hbar^{-1}\partial_k \xi$ and $\mathcal{A}^*_j=-\frac{eB\Omega}{\hbar}\frac{\hbar v_F k}{M} i\bra{n}\ket{\partial_j m}$ is the $j$ component of the modified Berry potential. By computing $g_{ij}$ and $\alpha_{kl}$ for the positive energy eigenstate
\begin{equation*}
    g_{xx}=\frac{1}{4} \frac{k_y^2}{k^4}\frac{\hbar^2v_F^2k^2}{\xi^2}+\Omega^2 \frac{\xi^2k_x^2}{\hbar^2v_F^2k^2}
\end{equation*}
\begin{equation*}
    g_{yy}=\frac{1}{4} \frac{k_x^2}{k^4}\frac{\hbar^2v_F^2k^2}{\xi^2}+\Omega^2 \frac{\xi^2k_y^2}{\hbar^2v_F^2k^2}
\end{equation*}
\begin{equation*}
    g_{xy}=g_{yx}=-\frac{1}{4} \frac{k_xk_y}{k^4}\frac{\hbar^2v_F^2k^2}{\xi^2}+\Omega^2 \frac{\xi^2k_xk_y}{\hbar^2v_F^2k^2}
\end{equation*}
\begin{equation*}
    \alpha_{kl}=\delta_{kl}\frac{v_F^2}{\xi}-\frac{\hbar^2v_F^4k_kk_l}{\xi^3}
\end{equation*}
it is worthy to show that actually, the third and fourth terms cancel and only the one coming from the corrections to the Berry potential holds, recovering the energy dispersion presented in Eq.\;\eqref{energy}. In this way, we can reach the grand potential $F$ determining the transport properties of the TIs in presence of perpendicular magnetic fields
\begin{equation}
    F=-k_BT \int \frac{d^2k}{(2\pi)^2} \left(1+\frac{eB\Omega^*}{\hbar}\right) ln(1+e^{-( \bar{\xi}-\mu)/k_BT})
\end{equation}
which incorporates the modified density of states and energy obtained with the changes of the Berry curvature and orbital magnetic moment. From here, we can compute the different transport magnitudes and coefficients such as, for instance, the system orbital magnetization $\mathcal{M}$ and susceptibility $\chi$. Thus, for $\mu=0$ and zero temperature, it is immediate to obtain the dependency of $\mathcal{M}$ with the external magnetic field $B$
\begin{equation}
    \mathcal{M}=-\frac{e^2v_F^2}{6\pi \abs{M}}B-\frac{3e^3\hbar v_F^4}{128\pi M^3} B^2 +\frac{e^4\hbar^2 v_F^6}{1260\pi \abs{M}^5} B^3
\end{equation}
and the orbital magnetic susceptibility $\chi=-(\partial^2 F/\partial B^2)$ with no more ingredients as their band gap $2M$ and Fermi velocity. Remarkably, we find a diamagnetic zero field susceptibility $\chi=-e^2v_F^2/(6\pi \abs{M})$, which is identical to that obtained in ref.\;\cite{PhysRevB.91.214405} ($\chi/\chi_0=-9\pi^2t/(6\pi\abs{M})$ with $t$ the first-neighbor hopping parameter), plus additional $B$-dependent terms which are not negligible for systems with small $M$. Notice that for zero gap systems ($M=0$) these corrections are not well-defined since the Berry curvature vanishes.

One outstanding application behind this relativistic formalism is its implementation to study thermoelectric features of TIs. In this case, it might not be desirable to introduce magnetic fields given that they break the time-reversal symmetry necessary for the preservation of Kramer\textquotesingle s pairs, which are responsible for their high efficient thermoelectric response \cite{Baldomir2019,PhysRevB.81.161302}. Notice that the Hamiltonian of a 2DTI is formed by two time-reversal copies of Hamiltonian (2), and introducing the same field on both non-interacting systems implies the breakdown of temporal invariance. However, there exists an equivalent form to introduce these interactions in a 4x4 Dirac Hamiltonian without breaking time-reversal symmetry. That way is the Dirac oscillator Hamiltonian    $H= M(\boldsymbol{k}) \beta + \boldsymbol{\alpha} \cdot (\boldsymbol{p}-im\omega \boldsymbol{r} \beta)$  \cite{moshinsky1989dirac,PhysRevA.76.041801},
which in essence incorporates a magnetic field $B=2m\omega/e$ with opposite signs on each one of the two time-reversal symmetry-related Hamiltonians given by Eq.\,\eqref{Hamiltonian} and its time-reversal counterpart $H'=\hat{T} H(k) \hat{T}^{-1}$, being $\hat{T}$ the time-reversal symmetry operator  \cite{PhysRevB.81.115407}. The Dirac oscillator is a powerful tool to examine relativistic interactions between electrons and chiral photons or thermal excitations in TIs \cite{PhysRevA.77.063815,PhysRevA.76.041801,failde2020emergent}. Besides the possibility to study higher-order effects, we have shown that these processes are compatible with the preservation of the topology and time-reversal symmetry, implying for the transport that at low fields the electric $\sigma=e^2/h \,(C-C')$ and electronic thermal conductivities $\kappa_e=\pi k_B^2/(6\hbar)\;(C-C')$ can remain quantized, being $C$ and $C'$ the Chern number of $H$ and $H'$ respectively \cite{PhysRevB.85.184503}.  Maintaining and combining these values with a good Seebeck coefficient and a low lattice thermal conductivity is determinant to obtain higher efficient thermoelectric devices \cite{Baldomir2019,PhysRevB.81.161302,PhysRevLett.112.226801}.

In summary, we provide a relativistic quantum derivation for two-dimensional topological systems with non-zero Berry curvature in presence of a perpendicular magnetic field. We have found that the change in the velocity of the electrons  due to the coupling of the magnetic field and the Berry curvature involves new corrections in their energy and magnetic moment which is associated with their relativistic nature. This is accompanied by a modulation of the Berry curvature that keeps the Chern number of the system invariant opening the door to study higher-order non-trivial magnetic and thermoelectric effects in Chern and topological insulators.

Authors acknowledge to CESGA, AEMAT ED431E 2018/08, PID2019-104150RB-I00 and the MAT2016-80762-R projects for financial support.

\providecommand{\noopsort}[1]{}\providecommand{\singleletter}[1]{#1}

\end{document}